\documentclass{jpsj2}
\title{ 
Momentum Dependent Local-Ansatz Approach to Correlated Electron 
Systems: Non Half-Filled Case
}

\author{
M. Atiqur R. \textsc{Patoary}\thanks{E-mail address:
k108609@eve.u-ryukyu.ac.jp}  and  Yoshiro \textsc{Kakehashi}\thanks{yok@sci.u-ryukyu.ac.jp}
}

\inst{
Department of Physics and Earth Sciences,
Faculty of Science, \\
University of the Ryukyus, \\
1 Senbaru, Nishihara, Okinawa, 903-0213, Japan
}

\abst{Momentum dependent local-ansatz wavefunction approach (MLA) to the correlated electron systems in solids has been further  developed to solve best a self-consistent equation for variational parameters at non half-filling. With use of the improved variational scheme we performed the numerical calculations for the non-half-filled band  Hubbard model on the hypercubic lattice in infinite dimensions. We verified that the self-consistent scheme significantly improves the correlation energy and the momentum distribution as compared with the original scheme in the MLA. We also demonstrate that the theory improves the standard variational methods such as the Local-Ansatz approach (LA) and the Gutzwiller wavefunction approach (GA); the ground-state energy in the MLA is lower than those of the LA and the GA in the weak and intermediate Coulomb interaction regimes. The double occupation number is shown to be suppressed as compared with the LA. Calculated momentum distribution functions show a distinct momentum dependence, which is qualitatively different from those of the LA and the GA.
}

\kword{variational method, electron correlations, Gutzwiller
wavefunction, local ansatz, Hubbard model, critical
Coulomb interaction, infinite dimensions}

\begin{document}
\maketitle

\section{Introduction} 
Electron correlations play an important role for understanding
the electronic structure, metal-insulator transition, and the 
high-temperature superconductivity in the solid-state physics. 
Thus many theories have been proposed so far to describe correlated 
electron system~\cite{fulde95,kakeh04-II} on the basis of the variational 
method~\cite{kakeh85,kakeh87,kakeh88},
the Green function techniques, as well as many numerical techniques such as 
the Monte-Carlo method~\cite{kakeh92,jarr92,kurt10}.

The variational theory among various methods has been developed as a practical method 
for understanding the ground-state properties of correlated 
electrons from  molecules to  solids over 50 years. 
 A minimum basis set to describe
correlated electrons is constructed in this approach by applying one-particle, two-particle,
and higher-order particle operators onto the Hartree-Fock wavefunction,
and their amplitudes are chosen to be best on the basis of
the variational principle. The Gutzwiller wavefunction  is one 
of the popular trial wavefunction in solids because of its simple 
and intuitive structure.  This approach was first introduced by Gutzwiller 
to clarify the role of electron correlations in metallic
ferromagnetism~\cite{gutz63,gutz64,gutz65}. There, one reduces the amplitudes 
of doubly occupied states on the local orbitals in the Hartree-Fock 
wavefunction by making use of a projection
operator $\Pi_{i} (1-gn_{i\uparrow}n_{i\downarrow})$.  Here
$n_{i\sigma}$ is the number operator for electrons on site $i$ with 
spin $\sigma$.  The variational parameter $g$ is determined by 
minimization of the ground-state energy. Brinkman and Rice  recognized 
that the Gutzwiller approximation describes the metal-insulator 
transition~\cite{br70}.  Because the Gutzwiller method is a
nonperturbative approach, it has extensively been applied to the strongly
correlated electron systems~\cite{geb97}.

The Gutzwiller wavefunction in the Gutzwiller ansatz approach (GA) 
yields a physical picture of electron correlations and is useful for correlation problems,
but it was not so easy to apply the method to realistic Hamiltonians.
The approach was successfully generalized by Stollhoff and Fulde
~\cite{stoll77,stoll78,stoll80} by using an alternative method called the
local-ansatz approach (LA), which is simpler in treatment and applicable 
to realistic Hamiltonians. The LA takes into account the excited states
created by local two-particle operators such as $\{ O_{i} \}=\{ \delta
n_{i\uparrow}\delta n_{i\downarrow} \}$, and determines their amplitudes
variationally.   Here $\delta n_{i\sigma}=
n_{i\sigma} - \langle n_{i\sigma} \rangle_{0}$, $\langle n_{i\sigma}
\rangle_{0}$ being the average electron number on site $i$ with spin
$\sigma$ in the Hartree-Fock approximation.  The theory has been applied to
many systems such as molecules, transition metals, polyacetylene, transition metal oxides 
and semiconductors~\cite{fulde95, fulde02}.  

Although the LA is able to explain fruitfully the  correlation effects
in actual materials, the application has been limited to the weakly 
correlated region because of the difficulty in evaluation 
of the higher-order terms in average quantities. The expansion of the Hilbert space by the local operators  is not sufficient  to characterize  precisely the weakly correlated states; the LA does not reduce to the second-order 
perturbation theory in the weak correlation limit. For the  Gutzwiller wavefunction, the same  difficulty   also arises even in infinite dimensions. 
To overcome the difficulty, Kakehashi $et.$  $al.$~\cite{kakeh08} proposed a variational wavefunction theory called the momentum-dependent local ansatz approach (MLA).

 When we expand the local operators
$\{O_i\}$ in the LA by means of the two particle operators in the momentum space, we find that each coefficient of the expansion is momentum independent. 
In the MLA wavefunction~\cite{kakeh08}, we replace the constant coefficients in the LA  with the momentum-dependent variational parameters in order to obtain the best local operators. It  results in a new set of local operators $\{\tilde O_i\}$. We then construct the MLA wavefunction with use of the local operators 
$\{\tilde O_i\}$ as $| \Psi_{\rm MLA} \rangle = \prod_i (1-\tilde{O_i}) | \phi_0 \rangle$. Here $| \phi_0 \rangle$ is the Hartree-Fock wavefunction and $i$ denotes site of atoms. The best local basis set is chosen by controlling the variational parameters in the momentum space. We calculate the ground-state energy using the MLA wavefunction within a single-site approximation (SSA). Minimizing the energy, 
we obtain a self-consistent equation with variational parameters. It is however difficult to solve the self-consistent equation directly. Because of this, we obtained in the previous paper, which we refer to I, an approximate solution which interpolates between the weak Coulomb interaction limit and the atomic limit.

In this paper we point out that it is indispensable toward quantitative calculations to choose the variational parameters best, though in our  paper I~\cite{kakeh08} we applied approximate variational parameters, and improve the variational parameters on the basis of the variational principle. We investigates the validity of our theory for the non half-filled case performing numerical calculations of  various physical quantities. Especially for non half-filled case we observe  that the best choice of variational parameters gives reasonable results for the momentum distribution, while the previous version of the variational parameters yields unphysical results near the Fermi level. Moreover, we demonstrate that the improved variational parameters  much improve the LA in the weak and intermediate correlation regimes.  

The outline of the paper is as follows. In the following section we write down our wavefunction for the single-band Hubbard model. We obtain  the ground-state energy within the SSA and derive the 
self-consistent equation for the momentum dependent variational parameters on the basis of the variational principle. We develop the theory to obtain the best value of variational parameters to solve the self-consistent equation. In \S 3, we present our results of  numerical calculations. We will clarify the role of the 
best choice of variational parameters on various  quantities. Furthermore, we discuss the correlation energy, the double occupation number, the momentum 
distribution function, and the quasiparticle weight as a function of 
the Coulomb interaction energy parameter, and  verify that the present approach
improves the LA in the weak and intermediate Coulomb interaction regimes for non half-filled band. We summarize our results in the last section and discuss the remaining problems.

\section{Momentum-Dependent Local Ansatz with the best Variational Parameters}
We adopt in this paper the single-band Hubbard model\cite{kakeh08,hub-I,hub-II,hub-III,kana63}  as
follows. 
\begin{eqnarray}
H = \sum_{i \sigma} (\epsilon_{0}-\sigma h) n_{i\sigma} 
+ \sum_{ij \sigma} t_{i j} \, a_{i \sigma}^{\dagger} a_{j \sigma} 
+ U \sum_{i} \, n_{i \uparrow} n_{i \downarrow} \ .
\label{hub}
\end{eqnarray}
Here $\epsilon_{0}$ ($h$) is the atomic level (magnetic field), 
$t_{ij}$ is the transfer integral between sites $i$ and $j$.  $U$
is the intra-atomic Coulomb energy parameter.  $ a_{i \sigma}^{\dagger}$
($ a_{i \sigma}$) denotes the creation (annihilation) operator for an
electron on site $i$ with spin $\sigma$, and 
$n_{i\sigma}=a_{i\sigma}^{\dagger} a_{i \sigma}$ is the electron density
operator on site $i$ for spin $\sigma$.

In the Hartree-Fock approximation, we neglect the fluctuation term  and replace the many-body Hamiltonian
(\ref{hub}) with an effective Hamiltonian for independent particle system,
\begin{eqnarray}
H_{0} = \sum_{ij \sigma} t_{i j\sigma} \, a_{i \sigma}^{\dagger} a_{j \sigma} 
- U \sum_{i} \, \langle n_{i \uparrow} \rangle_{0} 
\langle n_{i\downarrow} \rangle_{0} \ ,
\label{hf}
\end{eqnarray}
and approximate the ground-state wavefunction $|\Psi \rangle$ with that
of the Hartree-Fock Hamiltonian $H_{0}$, {\it i.e.}, 
$|\phi_{0} \rangle$.
Here $t_{ij\sigma}=(\epsilon_{0}+U\langle n_{i
-\sigma} \rangle_{0} - \sigma h)\delta_{ij} + t_{ij}(1-\delta_{ij})$. 
$\langle \sim \rangle_{0}$ denotes the Hartree-Fock average 
$\langle \phi_{0}| (\sim) |\phi_{0}\rangle$, and $\langle n_{i
\sigma} \rangle_{0}$ is the average electron number on site $i$ with
spin $\sigma$.
The original Hamiltonian (\ref{hub}) is then expressed as a sum of the
Hartree-Fock Hamiltonian (\ref{hf}) and the residual interactions as follows.
\begin{eqnarray}
H = H_{0} + U \sum_{i} \, O_{i} \ .
\label{hub2}
\end{eqnarray}
Here $O_{i}=\delta n_{i \uparrow}\delta n_{i \downarrow} $ and 
$\delta n_{i\sigma} = n_{i\sigma} - \langle n_{i\sigma} \rangle_{0}$.

In the  LA~\cite{stoll80}, the Hilbert space expanded by the local operators such as the residual Coulomb interactions $\{O_i\}=\delta n_{i\downarrow}\delta n_{i\uparrow}$ is taken into account in order to describe  the weak Coulomb interaction regime. The ansatz for the Hubbard model is written as 
\begin{eqnarray}
|\Psi_{\rm LA}\rangle = \Big[ \prod_{i} (1 - \eta^{}_{\rm \, LA} O_{i})
		    \Big]|\phi_{0} \rangle \ .
\label{lawf}
\end{eqnarray}
Here  $\eta^{}_{\rm \, LA}$ is the variational 
parameter as the amplitudes of the basis set expanded by $\{O_i\}$.

The LA is different from the Gutzwiller ansatz wavefunction 
$|\Psi_{\rm GA}\rangle = \Big[ \prod_{i} (1 - g n_{i \uparrow} n_{i \downarrow})
		    \Big]|\phi_{0}\rangle$
in which the doubly occupied states are explicitly controlled by a
variational parameter $g$ ($0 \le g \le 1 $), and simplify the evaluation of the physical quantities in the weakly correlated region.

As we have emphasized in our previous paper I~\cite{kakeh08},
though the LA is useful  for understanding correlation effects
in real system, the Hilbert space expanded by the  local operators $\{O_i\}$ is not sufficient to characterize exactly the weakly correlated region; it does not reduce to the second-order perturbation theory. In order to describe  the weak Coulomb interaction regime correctly we introduced  a new set of local operator 
\begin{eqnarray}
\tilde{O}_{i} = \sum_{k_{1}k_{2}k^{\prime}_{1}k^{\prime}_{2}} 
\langle k^{\prime}_{1}|i \rangle \langle i|k_{1} \rangle 
\langle k^{\prime}_{2}|i \rangle \langle i|k_{2} \rangle
\eta_{k^{\prime}_{2}k_{2}k^{\prime}_{1}k_{1}} 
\delta(a^{\dagger}_{k^{\prime}_{2}\downarrow}a_{k_{2}\downarrow})
\delta(a^{\dagger}_{k^{\prime}_{1}\uparrow}a_{k_{1}\uparrow}) \ , 
\label{otilde}
\end{eqnarray}
and  proposed the following new wavefunction with momentum dependent variational parameters  
$\{ \eta_{k^{\prime}_{2}k_{2}k^{\prime}_{1}k_{1}} \}$.
\begin{eqnarray}
|\Psi\rangle = \prod_{i} (1 - \tilde{O}_{i})|\phi_{0}\rangle \ .
\label{mla}
\end{eqnarray}
Here $\langle i|k \rangle = \exp (-i\boldsymbol{k} \cdot 
\boldsymbol{R}_{i}) / \sqrt{N}$ 
is an
overlap integral between the localized orbital and the Bloch state with
momentum $\boldsymbol{k}$, $\boldsymbol{R}_{i}$ denotes the atomic
position, and $N$ is the number of sites.
$ a_{k \sigma}^{\dagger}$ ($ a_{k \sigma}$) denotes the creation 
(annihilation) operator for an electron with momentum $\boldsymbol{k}$ 
and spin $\sigma$, and 
$\delta(a^{\dagger}_{k^{\prime}\sigma}a_{k\sigma})= 
a^{\dagger}_{k^{\prime}\sigma}a_{k\sigma} - \langle
a^{\dagger}_{k^{\prime}\sigma}a_{k\sigma} \rangle_{0}$.

The operator $\tilde{O}_{i}$ is still localized on site $i$
because of the projection $\langle k^{\prime}_{1}|i \rangle 
\langle i|k_{1} \rangle \langle k^{\prime}_{2}|i \rangle 
\langle i|k_{2} \rangle$.  
It should be noted that  
$\tilde{O}^{\dagger}_{i} \ne \tilde{O}_{i}$ and 
$\tilde{O}_{i}\tilde{O}_{j} \ne \tilde{O}_{j}\tilde{O}_{i}$
($i \ne j$) in general.
The wavefunction $|\Psi\rangle$ reduces to $|\Psi_{\rm LA} \rangle$ when 
$\{ \eta_{k^{\prime}_{2}k_{2}k^{\prime}_{1}k_{1}} \}$ become
momentum-independent because $\tilde{O_i}\rightarrow\eta_{\rm LA}O_i$ when $\eta_{k^{\prime}_{2}k_{2}k^{\prime}_{1}k_{1}}\rightarrow \eta_{\rm LA}$.

The variational parameters are determined by minimizing the ground-state
correlation energy $E_{\rm c}$. 
\begin{eqnarray}
E_{\rm c} = \langle H \rangle -  \langle H \rangle_{0} = 
\dfrac{\langle \Psi |\tilde{H}| \Psi \rangle}{\langle \Psi | \Psi \rangle} \ .
\label{corr}
\end{eqnarray}
Here $\tilde{H}=H - \langle H \rangle_{0}$.

Although it is not easy to calculate the correlation energy with use of 
the wavefunction (\ref{mla}), one can obtain the  
energy within the  single-site approximation (SSA). In the SSA, the average of $\langle \tilde{A} \rangle$  of an operator $\tilde{A}= A-\langle A \rangle_0$  with respect to the  wavefunction (\ref{mla}) is given as follows:
\begin{eqnarray}
\langle \tilde{A} \, \rangle = 
\sum_{i} \dfrac{\langle (1 - \tilde{O}^{\dagger}_{i}) \tilde{A}
(1 - \tilde{O}_{i}) \rangle_{0}}
{\langle (1 - \tilde{O}^{\dagger}_{i})(1 - \tilde{O}_{i}) \rangle_{0}} \ .
\label{ava}
\end{eqnarray}
The derivation of the above formula has been given in Appendix A of our paper I~\cite{kakeh08}.

By making use of the above formula, one can obtain the correlation
energy per atom as follows. 
\begin{eqnarray}
\epsilon_{\rm c} = \dfrac{-\langle
 \tilde{O}^{\dagger}_{i}\tilde{H}\rangle_{0} -
\langle \tilde{H} \tilde{O}_{i} \rangle_{0} + 
\langle \tilde{O}^{\dagger}_{i}\tilde{H}\tilde{O}_{i}\rangle_{0}}
{1 + \langle \tilde{O}^{\dagger}_{i}\tilde{O}_{i} \rangle_{0}} \ .
\label{ec}
\end{eqnarray}
Here we assumed that all the sites are equivalent to each other 
for simplicity and we
made use of the fact $\langle \tilde{O}^{\dagger}_{i} \rangle_{0}
= \langle \tilde{O}_{i} \rangle_{0} = 0$.

Each term in the correlation energy (\ref{ec}) can be calculated by
making use of Wick's theorem as follows.
\begin{eqnarray}
\langle \tilde{H} \tilde{O}_{i} \rangle_{0} & = &  U 
\sum_{k_{1}k_{2}k^{\prime}_{1}k^{\prime}_{2}} 
\langle k^{\prime}_{1}|i \rangle \langle i|k_{1} \rangle 
\langle k^{\prime}_{2}|i \rangle \langle i|k_{2} \rangle
\sum_{j}
\langle k_{1}|j \rangle \langle j|k^{\prime}_{1} \rangle 
\langle k_{2}|j \rangle \langle j|k^{\prime}_{2} \rangle 
\hspace{10mm}  \nonumber \\ 
& & \hspace*{15mm}  \times
\eta_{k^{\prime}_{2}k_{2}k^{\prime}_{1}k_{1}}
f(\tilde{\epsilon}_{k_{1}\uparrow})
(1-f(\tilde{\epsilon}_{k^{\prime}_{1}\uparrow}))
f(\tilde{\epsilon}_{k_{2}\downarrow})
(1-f(\tilde{\epsilon}_{k^{\prime}_{2}\downarrow})) \ ,
\label{ho}
\end{eqnarray}
\begin{eqnarray}
\langle \tilde{O}^{\dagger}_{i}\tilde{H} \rangle_{0} =
\langle \tilde{H} \tilde{O}_{i} \rangle^{\ast}_{0} \ ,
\hspace{90mm}
\label{oh}
\end{eqnarray}
\begin{eqnarray}
\langle \tilde{O}^{\dagger}_{i}\tilde{H}\tilde{O}_{i}\rangle_{0} & = & 
\sum_{k_{1}k_{2}k^{\prime}_{1}k^{\prime}_{2}} 
\langle i|k^{\prime}_{1} \rangle \langle k_{1}|i \rangle
\langle i|k^{\prime}_{2} \rangle \langle k_{2}|i \rangle \,
\eta^{\ast}_{k^{\prime}_{2}k_{2}k^{\prime}_{1}k_{1}}  \nonumber \\
& & \hspace*{-20mm} \times
f(\tilde{\epsilon}_{k_{1}\uparrow})
(1-f(\tilde{\epsilon}_{k^{\prime}_{1}\uparrow}))
f(\tilde{\epsilon}_{k_{2}\downarrow})
(1-f(\tilde{\epsilon}_{k^{\prime}_{2}\downarrow})) 
\sum_{k_{3}k_{4}k^{\prime}_{3}k^{\prime}_{4}} 
\langle k^{\prime}_{3}|i \rangle \langle i|k_{3} \rangle 
\langle k^{\prime}_{4}|i \rangle \langle i|k_{4} \rangle  \nonumber \\
& & \times
\left( \Delta E_{k^{\prime}_{2}k_{2}k^{\prime}_{1}k_{1}}
\delta_{k_{1}k_{3}}\delta_{k^{\prime}_{1}k^{\prime}_{3}}
\delta_{k_{2}k_{4}}\delta_{k^{\prime}_{2}k^{\prime}_{4}}
+ U_{k^{\prime}_{2}k_{2}k^{\prime}_{1}k_{1}
k^{\prime}_{4}k_{4}k^{\prime}_{3}k_{3}}
\right)
\eta_{k^{\prime}_{4}k_{4}k^{\prime}_{3}k_{3}} \ ,
\label{oho}
\end{eqnarray}
\begin{eqnarray}
U_{k^{\prime}_{2}k_{2}k^{\prime}_{1}k_{1}
k^{\prime}_{4}k_{4}k^{\prime}_{3}k_{3}} & = & 
U \sum_{j} [
\langle j|k_{1} \rangle \langle k_{3}|j \rangle 
f(\tilde{\epsilon}_{k_{3}\uparrow})\delta_{k^{\prime}_{1}k^{\prime}_{3}}
- \langle k^{\prime}_{1}|j \rangle \langle j|k^{\prime}_{3} \rangle
(1 - f(\tilde{\epsilon}_{k^{\prime}_{3}\uparrow})) \delta_{k_{1}k_{3}}
]   \nonumber \\ 
& & \times
[\langle j|k_{2} \rangle \langle k_{4}|j \rangle 
f(\tilde{\epsilon}_{k_{4}\downarrow})\delta_{k^{\prime}_{2}k^{\prime}_{4}}
- \langle k^{\prime}_{2}|j \rangle \langle j|k^{\prime}_{4} \rangle
(1 - f(\tilde{\epsilon}_{k^{\prime}_{4}\downarrow})) \delta_{k_{2}k_{4}}
] \ ,
\label{uk}
\end{eqnarray}
\begin{eqnarray}
\langle \tilde{O}^{\dagger}_{i}\tilde{O}_{i} \rangle_{0} & = &
\dfrac{1}{N^{4}} \sum_{k_{1}k_{2}k^{\prime}_{1}k^{\prime}_{2}}
|\eta_{k^{\prime}_{2}k_{2}k^{\prime}_{1}k_{1}}|^{2}
f(\tilde{\epsilon}_{k_{1}\uparrow})
(1-f(\tilde{\epsilon}_{k^{\prime}_{1}\uparrow}))
f(\tilde{\epsilon}_{k_{2}\downarrow})
(1-f(\tilde{\epsilon}_{k^{\prime}_{2}\downarrow})) \ . 
\label{oo}
\end{eqnarray}
Here $\tilde{\epsilon}_{k\sigma}=\epsilon_{k\sigma}-\mu$,
$\epsilon_{k\sigma}=\epsilon_{0}+U\langle n_{i -\sigma} \rangle_{0} +
\epsilon_{k} - \sigma h$ and  
 $ \Delta E_{k^{\prime}_{2}k_{2}k^{\prime}_{1}k_{1}} = 
\epsilon_{k^{\prime}_{2}\downarrow} - \epsilon_{k_{2}\downarrow}
+ \epsilon_{k^{\prime}_{1}\uparrow} - \epsilon_{k_{1}\uparrow}$
is a two-particle excitation energy. $\epsilon_k$ is the the Fourier transform of $t_{ij}$.

The above expressions (\ref{ho}) and (\ref{uk}) contain nonlocal terms 
via summation over $j$ ({\it i.e.}, $\sum_{j}$).
We thus make additional SSA called the $R=0$
approximation~\cite{kajz78,treglia80}.  
For example, we have in eq. (\ref{ho}) 
\begin{eqnarray}
\sum_{j}
\langle k^{\prime}_{1}|i \rangle \langle i|k_{1} \rangle 
\langle k^{\prime}_{2}|i \rangle \langle i|k_{2} \rangle
\langle k_{1}|j \rangle \langle j|k^{\prime}_{1} \rangle 
\langle k_{2}|j \rangle \langle j|k^{\prime}_{2} \rangle 
= \dfrac{1}{N^{4}} \sum_{j} 
{\rm e}^{i(k_{1}+k_{2}-k^{\prime}_{1}-k^{\prime}_{2})(R_{j}-R_{i})} \ ,
\label{r0}
\end{eqnarray}
but we only take into account the local term ($j=i$).
In the $R=0$ approximation,  
$\langle \tilde{H} \tilde{O}_{i}\rangle_{0} 
(= \langle \tilde{O}^{\dagger}_{i}\tilde{H} \rangle_{0}^{\ast})$,
and 
$\langle \tilde{O}^{\dagger}_{i}\tilde{H}\tilde{O}_{i}\rangle_{0}$
reduce as follows. 
\begin{eqnarray}
\langle \tilde{H} \tilde{O}_{i} \rangle_{0} & = &  \frac{U}{N^{4}} 
\sum_{k_{1}k_{2}k^{\prime}_{1}k^{\prime}_{2}} 
f(\tilde{\epsilon}_{k_{1}\uparrow})
(1-f(\tilde{\epsilon}_{k^{\prime}_{1}\uparrow}))
f(\tilde{\epsilon}_{k_{2}\downarrow})
(1-f(\tilde{\epsilon}_{k^{\prime}_{2}\downarrow})) \,
\eta_{k^{\prime}_{2}k_{2}k^{\prime}_{1}k_{1}} \ ,
\label{hor0}
\end{eqnarray}
\begin{eqnarray}
\langle \tilde{O}^{\dagger}_{i}\tilde{H}\tilde{O}_{i}\rangle_{0} & = & 
\dfrac{1}{N^{4}} \sum_{k_{1}k_{2}k^{\prime}_{1}k^{\prime}_{2}} 
f(\tilde{\epsilon}_{k_{1}\uparrow})
(1-f(\tilde{\epsilon}_{k^{\prime}_{1}\uparrow}))
f(\tilde{\epsilon}_{k_{2}\downarrow})
(1-f(\tilde{\epsilon}_{k^{\prime}_{2}\downarrow})) \,
\eta^{\ast}_{k^{\prime}_{2}k_{2}k^{\prime}_{1}k_{1}}  
\hspace{20mm}  \nonumber \\
& &  \times
\bigg[ 
\Delta E_{k^{\prime}_{2}k_{2}k^{\prime}_{1}k_{1}} \,
\eta_{k^{\prime}_{2}k_{2}k^{\prime}_{1}k_{1}}  \nonumber \\
& &  \hspace{-10mm}
+ \dfrac{U}{N^{2}}
\Big\{
\sum_{k_{3}k_{4}}
f(\tilde{\epsilon}_{k_{3}\uparrow})f(\tilde{\epsilon}_{k_{4}\downarrow}) \,
\eta_{k^{\prime}_{2}k_{4}k^{\prime}_{1}k_{3}}
- \sum_{k_{3}k^{\prime}_{4}}
f(\tilde{\epsilon}_{k_{3}\uparrow})
(1-f(\tilde{\epsilon}_{k^{\prime}_{4}\downarrow})) \,
\eta_{k^{\prime}_{4}k_{2}k^{\prime}_{1}k_{3}}  \nonumber \\
& & \hspace*{-20mm}
- \sum_{k^{\prime}_{3}k_{4}}
(1 - f(\tilde{\epsilon}_{k^{\prime}_{3}\uparrow}))
f(\tilde{\epsilon}_{k_{4}\downarrow}) \,
\eta_{k^{\prime}_{2}k_{4}k^{\prime}_{3}k_{1}}
+ \sum_{k^{\prime}_{3}k^{\prime}_{4}}
(1 - f(\tilde{\epsilon}_{k^{\prime}_{3}\uparrow}))
(1 - f(\tilde{\epsilon}_{k^{\prime}_{4}\downarrow})) \,
\eta_{k^{\prime}_{4}k_{2}k^{\prime}_{3}k_{1}}
\Big\}
\bigg] \ .
\label{ohor0}
\end{eqnarray}

Variational parameters 
$\{ \eta_{k^{\prime}_{2}k_{2}k^{\prime}_{1}k_{1}} \}$
are obtained by minimizing the correlation energy $\epsilon_{\rm c}$, 
{\it i.e.}, eq. (\ref{ec}) with eqs. (\ref{oo}), (\ref{hor0}), and
(\ref{ohor0}).
The self-consistent equations for 
$\{ \eta_{k^{\prime}_{2}k_{2}k^{\prime}_{1}k_{1}} \}$
in the SSA are given as follows.
\begin{eqnarray}
(\Delta E_{{k^{\prime}_{2}k_{2}k^{\prime}_{1}k_{1}}} - \epsilon_{\rm c})
\eta_{k^{\prime}_{2}k_{2}k^{\prime}_{1}k_{1}}  \hspace{5mm} \nonumber \\
& &  \hspace{-53mm}
+ \dfrac{U}{N^{2}}
\Big[
\sum_{k_{3}k_{4}}
f(\tilde{\epsilon}_{k_{3}\uparrow})f(\tilde{\epsilon}_{k_{4}\downarrow})
\eta_{k^{\prime}_{2}k_{4}k^{\prime}_{1}k_{3}}
- \sum_{k_{3}k^{\prime}_{4}}
f(\tilde{\epsilon}_{k_{3}\uparrow})
(1-f(\tilde{\epsilon}_{k^{\prime}_{4}\downarrow}))
\eta_{k^{\prime}_{4}k_{2}k^{\prime}_{1}k_{3}}  \nonumber \\
& & \hspace*{-53mm}
- \sum_{k^{\prime}_{3}k_{4}}
(1 - f(\tilde{\epsilon}_{k^{\prime}_{3}\uparrow}))
f(\tilde{\epsilon}_{k_{4}\downarrow})
\eta_{k^{\prime}_{2}k_{4}k^{\prime}_{3}k_{1}}
+ \sum_{k^{\prime}_{3}k^{\prime}_{4}}
(1 - f(\tilde{\epsilon}_{k^{\prime}_{3}\uparrow}))
(1 - f(\tilde{\epsilon}_{k^{\prime}_{4}\downarrow}))
\eta_{k^{\prime}_{4}k_{2}k^{\prime}_{3}k_{1}}
\Big] = U \, . \ \ \ \
\label{eqeta}
\end{eqnarray}

It should be noted that the variational parameters $\{\eta_{{k_2}^\prime k_2 {k_1}^\prime k_1 } \}$ in eq. (\ref{eqeta}) vanish when $U\longrightarrow 0$, $i.e.,$ ${\eta_{k'_2 k_2 k'_1 k_1 }} \sim O(U)$.
Thus in the weak $U$ limit, one can omit
the second term at the l.h.s. (left-hand-side).
We then obtain the solution in the weak $U$ limit as
\begin{eqnarray}
\eta_{k^{\prime}_{2}k_{2}k^{\prime}_{1}k_{1}} = 
\dfrac{U}{\Delta E_{{k^{\prime}_{2}k_{2}k^{\prime}_{1}k_{1}}}} \ .
\label{etaweak}
\end{eqnarray}

In the atomic limit the transfer integrals $t_{ij}$ disappear, and one electron energy eigen value $\epsilon_k$ becomes $k$-independent, $\epsilon_0$. Thus,  $\Delta E_{k'_2 k_2 k'_1 k_1 }$ vanishes. In this limit we can drop the $k$ dependence of  ${\eta_{k'_2 k_2 k'_1 k_1 }}$, i.e.,
${\eta_{k'_2 k_2 k'_1 k_1 }} \longrightarrow \eta$. 
Then, we find a $k$-independent solution being identical with the LA.
\begin{eqnarray}
\eta^{}_{\rm \, LA} = 
\frac{\displaystyle -\langle O_{i}\tilde{H}O_{i}\rangle_{0} + 
\sqrt{\langle O_{i}\tilde{H}O_{i}\rangle_{0}^{2}+
4\langle O_{i}\tilde{H}\rangle^{2}_{0}\langle O_{i}^{2}\rangle_{0}}
}
{2\langle O_{i}\tilde{H}\rangle_{0}\langle O_{i}^{2}\rangle_{0}
} \ .
\label{etala}
\end{eqnarray}

It is not easy to find the solution of eq. (\ref{eqeta}) for the
intermediate strength of Coulomb interaction $U$.  We therefore proposed in our paper I
an approximate solution which interpolates between the weak and the
atomic limits; we approximate 
$\{ \eta_{k^{\prime}_{2}k_{2}k^{\prime}_{1}k_{1}} \}$ in the second term
with the momentum-independent parameter $\eta$ which is suitable for the
atomic region.  We have then 
\begin{eqnarray}
\eta_{k^{\prime}_{2}k_{2}k^{\prime}_{1}k_{1}} = 
\dfrac{U[1 - \eta (1 - 2 \langle n_{i\uparrow} \rangle_{0}) 
(1 - 2 \langle n_{i\downarrow} \rangle_{0})]}
{\Delta E_{{k^{\prime}_{2}k_{2}k^{\prime}_{1}k_{1}}} - \epsilon_{\rm c}} \ .
\label{etaint}
\end{eqnarray}
In the previous paper I~\cite{kakeh08} we made use of that in the LA for $\eta$, and  adopted the correlation energy  in the LA for $\epsilon_c$ in the denominator. We call this the non-self-consistent MLA in the followings. 

The best value of $\eta$, however, should be determined variationally in general. 
In this paper we further develop  the theory in which $\eta$ is determined best. According to the variational principle, the ground-state energy 
$E_0$ satisfies the the following inequality.
\begin{equation}
E_0 \leq E[\Psi] =\frac{ \langle \Psi |H| \Psi \rangle }
{ \langle \Psi | \Psi \rangle } . 
\end{equation}
Here $\Psi$ is a trial wavefunction.

In the MLA, we choose the wavefunction $\Psi=\Psi(\{\eta_{k'_2 k_2 k'_1 k_1 }\})$ 
and the corresponding energy $E( \{ \eta_{k'_2 k_2 k'_1 k_1 }\})$ satisfies the inequality $ E_0  \leq E( \{ \eta_{k'_2 k_2 k'_1 k_1 }\})$.
For the stationary values $\eta_{k'_2 k_2 k'_1 k_1}^{\ast} $ , we have 
\begin{equation}
E_0 \leq E( \{ \eta_{k'_2 k_2 k'_1 k_1 }^{\ast}\})
\leq E( \{ \eta_{k'_2 k_2 k'_1 k_1 }\}) \ .
\end{equation} 

In the previous calculations\cite{kakeh08}, we obtained an approximate $\eta_{k'_2 k_2 k'_1 k_1 }^{\ast}$  (\ref{etaint});
\begin{eqnarray}
\eta_{k^{\prime}_{2}k_{2}k^{\prime}_{1}k_{1}} (\tilde{\eta} , \epsilon_c)= 
\dfrac{U\tilde{\eta}}
{\Delta E_{{k^{\prime}_{2}k_{2}k^{\prime}_{1}k_{1}}} - \epsilon_{\rm c}} \ ,
\label{etaint2}
\end{eqnarray}
which is determined by the correlation energy $\epsilon_c$ and a momentum independent parameter $\tilde\eta$.
Here
\begin {equation}
\tilde \eta = [1 - \eta (1 - 2 \langle n_{i\uparrow} \rangle_{0})
(1 - 2 \langle n_{i\downarrow} \rangle_{0})] .
\label{etatil0}
\end{equation}
When we adopt the form (\ref{etaint2}) as a trial set of amplitudes, we have an inequality as 
\begin{equation}
\langle E_0\rangle \leq E( \{ \eta_{k'_2 k_2 k'_1 k_1 }^{\ast}\})
\leq E( \{ \eta_{k'_2 k_2 k'_1 k_1 } (\tilde\eta , \epsilon_c)\}) \ .
\end{equation}
The above relation implies that  the best $\tilde{\eta}$ is again determined from the stationary 
condition of the trial energy $E( \{ \eta_{k'_2 k_2 k'_1 k_1 } (\tilde\eta , \epsilon_c)\})$. 
Because $\epsilon_c$ should satisfy the stationary condition 
$\delta \epsilon_c =0$ for the value $\tilde\eta^\ast$,  $\tilde\eta^\ast$ 
is determined by the following condition
\begin{equation}
\Bigg [\frac {\partial \epsilon ( \{ \eta_{k'_2 k_2 k'_1 k_1 } (\tilde\eta , \epsilon_c)\})}
{\partial \tilde\eta}\Bigg]_{\epsilon_c} \ = \ 0 \ .
\label{stac}
\end{equation}
The  self-consistent equation is obtained from eq.(\ref{stac}) in the same way as in eq. (\ref{eqeta})
\begin{align}
& \hspace{-6mm}\frac {1}{ N^4}\sum_{ k_1 k'_1 k_2 k'_2} 
f(\tilde{\epsilon}_{k_1 \uparrow}) [1- f(\tilde{\epsilon}_{k'_1\uparrow})]
f(\tilde{\epsilon}_{k_2 \downarrow}) [1- f(\tilde{\epsilon}_{k'_2 \downarrow})]
\frac{\partial {\eta_{k'_2 k_2 k'_1 k_1 }} }{\partial \tilde\eta}
(\Delta E_{k'_2 k_2 k'_1 k_1}-\epsilon_c){\eta_{k'_2 k_2 k'_1 k_1 }} \nonumber \\
& + \frac {U}{N^6}\sum_{ k_1 k'_1 k_2 k'_2} 
f(\tilde{\epsilon}_{k_1 \uparrow}) [1- f(\tilde{\epsilon}_{k'_1 \uparrow})]
f(\tilde{\epsilon}_{k_2 \downarrow}) [1- f(\tilde{\epsilon}_{k'_2 \downarrow})]
\frac{\partial {\eta_{k'_2 k_2 k'_1 k_1 }} }{\partial \tilde\eta} \nonumber \\
& \times \Big[  \sum_{k_3 k_4} f( \tilde{\epsilon}_{k_3 \uparrow}) 
f( \tilde{\epsilon}_{ k_4 \downarrow} ) \eta_{k'_2 k_4 k'_1 k_3 }
-\sum_{k'_3 k_4} [1-f( \tilde{\epsilon}_{k'_3\uparrow})] 
f( \tilde{\epsilon}_{ k_4 \downarrow} ) \eta_{k'_2 k_4 k'_3 k_1 }\nonumber \\
&-\sum_{k_3 k'_4} f( \tilde{\epsilon}_{k_3 \uparrow})
[1-f( \tilde{\epsilon}_{ k'_4 \downarrow} )] \eta_{k'_4 k_2 k'_1 k_3 } 
+ \sum_{k'_3 k'_4} [1-f(\tilde{\epsilon}_{k'_3 \uparrow})] [1-f(\tilde{\epsilon}_{{k'_4}\downarrow} ) ]
\eta_{k'_4 k_2 k'_3 k_1 } \Big] \nonumber \\
& = \frac { U}{N^4}\sum_{ k_1 k'_1 k_2 k'_2} 
f(\tilde{\epsilon}_{k_1 \uparrow}) [1- f(\tilde{\epsilon}_{k'_1 \uparrow})]
f(\tilde{\epsilon}_{k_2 \downarrow}) [1- f(\tilde{\epsilon}_{k'_2 \downarrow})]
\frac{\partial {\eta_{k'_2 k_2 k'_1 k_1 }} }{\partial \tilde\eta} . 
\label{sceq2}
\end{align} 
Here  ${\partial {\eta_{k'_2 k_2 k'_1 k_1 }} }/{\partial \tilde\eta}$
is obtained  from eq. (\ref{etaint2}) as
\begin{equation} 
\frac{\partial {\eta_{k'_2 k_2 k'_1 k_1 }} }{\partial \tilde\eta}= 
\dfrac{U}
{\Delta E_{{k^{\prime}_{2}k_{2}k^{\prime}_{1}k_{1}}} - \epsilon_c} \ .
\end {equation}
Substituting the above expression into  the  self-consistent 
equation (\ref{sceq2}) we obtain 
\begin{equation}
\tilde \eta=\dfrac{1}{1+\dfrac{U C}{D}} \ .
\label{etatil}
\end{equation}
Here
\begin{align}
C &= \frac {1}{N^6}\sum_{ k_1 k'_1 k_2 k'_2} 
\frac {f(\tilde{\epsilon}_{k_1 \uparrow}) [1- f(\tilde{\epsilon}_{k'_1\uparrow})]
f(\tilde{\epsilon}_{k_2 \downarrow}) [1- f(\tilde{\epsilon}_{k'_2 \downarrow})]}
{(\Delta E_{k'_2 k_2 k'_1 k_1}-\epsilon_c)} \nonumber \\
&\times \Big \{  \sum_{k_3 k_4} 
\frac{f( \tilde{\epsilon}_{k_3 \uparrow}) f( \tilde{\epsilon}_{ k_4 \downarrow} )} {(\Delta E_{k'_2 k_4 k'_1 k_3}-\epsilon_c)}
-\sum_{k'_3 k_4}\frac {[1-f( \tilde{\epsilon}_{k'_3\uparrow})] 
f( \tilde{\epsilon}_{ k_4 \downarrow} )}{(\Delta E_{k'_2 k_4 k'_3 k_1}-\epsilon_c)}\nonumber \\
&-\sum_{k_3 k'_4} \frac{f( \tilde{\epsilon}_{k_3 \uparrow})
[1-f( \tilde{\epsilon}_{ k'_4 \downarrow} )]} {(\Delta E_{k'_4 k_2 k'_1 k_3}-\epsilon_c)} 
+ \sum_{k'_3 k'_4} \frac{[1-f(\tilde{\epsilon}_{k'_3 \uparrow})] [1-f(\tilde{\epsilon}_{{k'_4}\downarrow} ) ]}
{(\Delta E_{k'_4 k_2 k'_3 k_1}-\epsilon_c)} \Big \},
\label{etatilde}
\end {align}

and 
\begin{equation}
D = \frac {1}{N^4}\sum_{ k_1 k'_1 k_2 k'_2} 
\frac {f(\tilde{\epsilon}_{k_1 \uparrow}) [1- f(\tilde{\epsilon}_{k'_1\uparrow})]
f(\tilde{\epsilon}_{k_2 \downarrow}) [1- f(\tilde{\epsilon}_{k'_2 \downarrow})]}
{(\Delta E_{k'_2 k_2 k'_1 k_1}-\epsilon_c)} \  .
\label{etatilnu}
\end {equation}

In the energy representation, each term is expressed as follows:
\begin{align}
C &=  \int \frac{\bigg[ \prod \limits^6_{n=1} d\epsilon_{n}\bigg] 
\bigg[ \prod \limits^6_{n=1} \rho{(\epsilon_n)}\bigg] 
f(\tilde{\epsilon}_{1\uparrow}) 
[1-f(\tilde{\epsilon}_{2\uparrow})] 
f(\tilde{\epsilon}_{3\downarrow})
[1-f(\tilde{\epsilon}_{4\downarrow})]
f(\tilde{\epsilon}_{5\uparrow})
f(\tilde{\epsilon}_{6\downarrow})}
{(\epsilon_{4} - \epsilon_{3} + \epsilon_{2} - \epsilon_{1} -\epsilon_c)
(\epsilon_{4} - \epsilon_{6} + \epsilon_{2} - \epsilon_{5} -\epsilon_c)} \nonumber  \\
& -\int \frac{\bigg[ \prod \limits^6_{n=1} d\epsilon_{n}\bigg] 
\bigg[ \prod \limits^6_{n=1} \rho{(\epsilon_n)}\bigg] 
f(\tilde{\epsilon}_{1\uparrow}) 
[1-f(\tilde{\epsilon}_{2\uparrow})] 
f(\tilde{\epsilon}_{3\downarrow})
[1-f(\tilde{\epsilon}_{4\downarrow})]
[1-f(\tilde{\epsilon}_{5\uparrow})]
f(\tilde{\epsilon}_{6\downarrow})}
{(\epsilon_{4} - \epsilon_{3} + \epsilon_{2} - \epsilon_{1} -\epsilon_c)
(\epsilon_{4} - \epsilon_{6} + \epsilon_{5} - \epsilon_{1} -\epsilon_c)} \nonumber  \\
& -\int \frac{\bigg[ \prod \limits^6_{n=1} d\epsilon_{n}\bigg] 
\bigg[ \prod \limits^6_{n=1} \rho{(\epsilon_n)}\bigg] 
f(\tilde{\epsilon}_{1\uparrow}) 
[1-f(\tilde{\epsilon}_{2\uparrow})] 
f(\tilde{\epsilon}_{3\downarrow})
[1-f(\tilde{\epsilon}_{4\downarrow})]
f(\tilde{\epsilon}_{5\uparrow})
[1-f(\tilde{\epsilon}_{6\downarrow})]}
{(\epsilon_{4} - \epsilon_{3} + \epsilon_{2} - \epsilon_{1} -\epsilon_c)
(\epsilon_{6} - \epsilon_{3} + \epsilon_{2} - \epsilon_{5} -\epsilon_c)} \nonumber  \\
& +\int \frac{\bigg[ \prod \limits^6_{n=1} d\epsilon_{n}\bigg] 
\bigg[ \prod \limits^6_{n=1} \rho{(\epsilon_n)}\bigg] 
f(\tilde{\epsilon}_{1\uparrow}) 
[1-f(\tilde{\epsilon}_{2\uparrow})] 
f(\tilde{\epsilon}_{3\downarrow})
[1-f(\tilde{\epsilon}_{4\downarrow})]
[1-f(\tilde{\epsilon}_{5\uparrow})]
[1-f(\tilde{\epsilon}_{6\downarrow})]}
{(\epsilon_{4} - \epsilon_{3} + \epsilon_{2} - \epsilon_{1} -\epsilon_c)
(\epsilon_{6} - \epsilon_{3} + \epsilon_{5} - \epsilon_{1} -\epsilon_c)}  ,
\label{etatlde}
%\end{equation} 
\end{align} 
%\normalsize
%
%
and 
\begin{equation} 
D = \int \frac{\bigg[ \prod \limits^4_{n=1} d\epsilon_{n}\bigg] 
\bigg[ \prod \limits^4_{n=1} \rho{(\epsilon_n)}\bigg] 
f(\tilde{\epsilon}_{1\uparrow}) 
[1-f(\tilde{\epsilon}_{2\uparrow})] 
f(\tilde{\epsilon}_{3\downarrow})
[1-f(\tilde{\epsilon}_{4\downarrow})]}
{\epsilon_{4} - \epsilon_{3} + \epsilon_{2} - \epsilon_{1} -\epsilon_c} \ .
\label{etatlnu}
\end{equation} 
Here $\tilde{\epsilon}_{n\sigma}=\epsilon_{n}+\tilde\epsilon_\sigma$ and  $\tilde {\epsilon}_\sigma = \epsilon_0+U \langle n_{-\sigma} \rangle_0 -\mu $
is the atomic level measured from the chemical potential, 
and $\rho {(\epsilon)}$  is the density of states for the one-electron energy eigen 
values  for the non-interacting system $t_{ij}$. 

It should be noted that the self-consistent solution (\ref{etatil}) is also obtained by solving approximately the original self-consistent 
eq.(\ref{eqeta}).  In order to do this, first we divide the both sides of eq.(\ref{eqeta}) by $(\Delta E_{k'_2 k_2 k'_1 k_1}-\epsilon_c)$ , substitute the form (\ref{etaint2}), and we obtain eq. (\ref{etatil}) after taking the average with respect to $k_1 k'_1 k_2 k'_2$ with a weight $f(\tilde{\epsilon}_{k_1 \uparrow}) [1- f(\tilde{\epsilon}_{k'_1\uparrow})]
f(\tilde{\epsilon}_{k_2 \downarrow}) [1- f(\tilde{\epsilon}_{k'_2 \downarrow})]$. The variational principles on $\tilde\eta$ tells  us that such a solution should be the best among possible approximate solutions. We also note  that an approximate form (\ref{etatil0}), which was obtained in paper I, is derived from the solution (\ref{etatil}). In fact,  we rewrite eq. (\ref{etatil}) as $\tilde{\eta}= 1-U\tilde{\eta}C/D$. By replacing the approximate form (\ref{etaint2})
in the expression of $U\tilde{\eta}C$ with the momentum independent value $\eta$, we reach eq. (\ref{etatil0}).

The ground-state correlation energy is obtained by substituting the variational parameters (\ref{etaint2}) into eq. (\ref{ec}).  The each element in the energy is given as follows.
\begin{align} 
&\langle \tilde{H} \tilde{O}_{i}\rangle_{0} =
\langle \tilde{O}^{\dagger}_{i}\tilde{H} \rangle_{0}^{\ast} \nonumber \\
&=  U^{2} \tilde \eta
\int \frac{\bigg[ \prod \limits^4_{n=1} d\epsilon_{n}\bigg] 
\bigg[ \prod \limits^4_{n=1} \rho{(\epsilon_n)}\bigg] 
f(\tilde{\epsilon}_{1\uparrow}) 
[1-f(\tilde{\epsilon}_{2\uparrow})] 
f(\tilde{\epsilon}_{3\downarrow})
[1-f(\tilde{\epsilon}_{4\downarrow})}
{\epsilon_{4} - \epsilon_{3} + \epsilon_{2} - \epsilon_{1} -\epsilon_c} \ ,
\label{ho2}
\end{align}
\begin{eqnarray}
\langle \tilde{O}^{\dagger}_{i} \tilde{H} \tilde{O}_{i} \rangle_{0} = 
\langle \tilde{O}^{\dagger}_{i} \tilde{H}_{0} \tilde{O}_{i} \rangle_{0} 
+ U \langle \tilde{O}^{\dagger}_{i} O_{i} \tilde{O}_{i} \rangle_{0} \ ,
\label{oho2}
\end{eqnarray}
\begin{align}
&\langle \tilde{O}^{\dagger}_{i} \tilde{H}_{0} \tilde{O}_{i} \rangle_{0} \nonumber \\
& =  U^{2} \tilde \eta^{2} 
\int \frac{\bigg[ \prod \limits^4_{n=1} d\epsilon_{n}\bigg] 
\bigg[ \prod \limits^4_{n=1} \rho{(\epsilon_n)}\bigg] 
f(\tilde{\epsilon}_{1\uparrow}) 
[1-f(\tilde{\epsilon}_{2\uparrow})] 
f(\tilde{\epsilon}_{3\downarrow})
[1-f(\tilde{\epsilon}_{4\downarrow})]}
{(\epsilon_{4} - \epsilon_{3} + \epsilon_{2} - \epsilon_{1} -\epsilon_c)^{2}
(\epsilon_{4} - \epsilon_{3} + \epsilon_{2} - \epsilon_{1})^{-1}} \,
\label{oh0o2}
\end{align}
\begin{align}
& \langle \tilde{O}^{\dagger}_{i} O_{i} \tilde{O}_{i} \rangle_{0} \nonumber \\
&= U^{2}\tilde \eta^2 
\int \frac{\bigg[ \prod \limits^4_{n=1} d\epsilon_{n}\bigg] 
\bigg[ \prod \limits^4_{n=1} \rho{(\epsilon_n)}\bigg] 
f(\tilde{\epsilon}_{1\uparrow}) 
[1-f(\tilde{\epsilon}_{2\uparrow})] 
f(\tilde{\epsilon}_{3\downarrow})
[1-f(\tilde{\epsilon}_{4\downarrow})}
{(\epsilon_{4} - \epsilon_{3} + \epsilon_{2} - \epsilon_{1} -\epsilon_c)^{2}} \nonumber \\
&\hspace*{20mm}\times \Bigg[\int \frac{\bigg[ \prod \limits^6_{n=5} d\epsilon_{n}\bigg] 
\bigg[ \prod \limits^6_{n=5} \rho{(\epsilon_n)}\bigg] 
f(\tilde{\epsilon}_{5\uparrow}) 
f(\tilde{\epsilon}_{6\downarrow})}
{(\epsilon_{4} - \epsilon_{6} + \epsilon_{2} - \epsilon_{5} -\epsilon_c)} \nonumber \\
&\hspace*{20mm}-\int \frac{\bigg[ \prod \limits^6_{n=5} d\epsilon_{n}\bigg] 
\bigg[ \prod \limits^6_{n=5} \rho{(\epsilon_n)}\bigg] 
f(\tilde{\epsilon}_{5\uparrow}) 
[1-f(\tilde{\epsilon}_{6\downarrow})]}
{(\epsilon_{6} - \epsilon_{3} + \epsilon_{2} - \epsilon_{5} -\epsilon_c)} \nonumber \\
&\hspace*{20mm}-\int \frac{\bigg[ \prod \limits^6_{n=5} d\epsilon_{n}\bigg] 
\bigg[ \prod \limits^6_{n=5} \rho{(\epsilon_n)}\bigg] 
[1-f(\tilde{\epsilon}_{5\uparrow})] 
f(\tilde{\epsilon}_{6\downarrow})}
{(\epsilon_{4} - \epsilon_{6} + \epsilon_{5} - \epsilon_{1} -\epsilon_c)} \nonumber \\
&\hspace*{20mm}+\int \frac{\bigg[ \prod \limits^6_{n=5} d\epsilon_{n}\bigg] 
\bigg[ \prod \limits^6_{n=5} \rho{(\epsilon_n)}\bigg] 
[1-f(\tilde{\epsilon}_{5\uparrow})] 
[1-f(\tilde{\epsilon}_{6\downarrow})]}
{(\epsilon_{6} - \epsilon_{3} + \epsilon_{5} - \epsilon_{1} -\epsilon_c)} \Bigg] \ ,
\label{ooo02}
\end{align}
\begin{align} 
\langle \tilde{O}^{\dagger}_{i}\tilde{O}_{i} \rangle_{0} 
=  U^{2} \tilde \eta^{2}
\int \frac{\bigg[ \prod \limits^4_{n=1} d\epsilon_{n}\bigg] 
\bigg[ \prod \limits^4_{n=1} \rho{(\epsilon_n)}\bigg] 
f(\tilde{\epsilon}_{1\uparrow}) 
[1-f(\tilde{\epsilon}_{2\uparrow})] 
f(\tilde{\epsilon}_{3\downarrow})
[1-f(\tilde{\epsilon}_{4\downarrow})}
{(\epsilon_{4} - \epsilon_{3} + \epsilon_{2} - \epsilon_{1} -\epsilon_c)^{2}}.
\label{oo2}
\end{align}

It should be noted that $\tilde{\eta}$ in eq. (\ref{etatil}) is given as a function of 
$\epsilon_c$, and $\epsilon_c$ in eq. (\ref{ec}) depends on $\tilde{\eta}$ and $\epsilon_c$. Therefore,
both equations have to be solved self-consistently. To determine the best value of $\tilde{\eta}$, we start from $\epsilon_c$ in the LA for example, and calculate $\tilde{\eta}$ according to eq. (\ref{etatil}). Next we calculate various elements $\langle \tilde{H} \tilde{O}_{i} \rangle_{0}$ 
($\langle \tilde{O}^{\dagger}_{i}\tilde{H} \rangle_{0}^{\ast}$),
$\langle \tilde{O}^{\dagger}_{i}\tilde{H}\tilde{O}_{i}\rangle_{0}$, and  
$\langle \tilde{O}^{\dagger}_{i}\tilde{O}_{i} \rangle_{0}$ which are  given by eqs. (\ref{ho2}), (\ref{oho2}), and  (\ref{oo2}), respectively. Using these values we calculate $\epsilon_c$ according to eq. (\ref{ec}).We repeat this cycle until the self-consistency of $\epsilon_c$ and 
$\tilde{\eta}$ is satisfied. We call this scheme the self-consistent MLA .

Electron number 
$\langle n_{i} \rangle (= \sum_{\sigma} \langle n_{i\sigma} \rangle)$,
the momentum distribution $\langle n_{k\sigma} \rangle$, and the 
double occupation number $\langle n_{i\uparrow}n_{i\downarrow} \rangle$
are obtained from $\partial \langle H \rangle / \partial \epsilon_0$, 
$\partial \langle H \rangle / \partial \hat{\epsilon}_{k\sigma}$, and
$\partial \langle H \rangle / \partial U_{i}$, 
respectively.  Here $\hat{\epsilon}_{k\sigma} = \epsilon_{k} - \sigma h$.
Making use of the single-site energy (\ref{ec}) and the Feynman-Hellmann
theorem~\cite{hell}, we obtain the following expressions.
\begin{eqnarray}
\langle n_{i} \rangle = \langle n_{i} \rangle_{0} + 
\dfrac{\langle \tilde{O}_{i} \tilde{n}_{i} \tilde{O}_{i} \rangle_{0}}
{1+\langle \tilde{O}^{\dagger}_{i} \tilde{O}_{i} \rangle_{0}} \ ,
\label{ni}
\end{eqnarray}
\begin{eqnarray}
\langle n_{k\sigma} \rangle = \langle n_{k\sigma} \rangle_{0} + 
\dfrac{N \langle \tilde{O}_{i} \tilde{n}_{k\sigma} \tilde{O}_{i} \rangle_{0}}
{1+\langle \tilde{O}^{\dagger}_{i} \tilde{O}_{i} \rangle_{0}} \ ,
\label{nk}
\end{eqnarray}
\begin{eqnarray}
\langle n_{i\uparrow}n_{i\downarrow} \rangle = 
\langle n_{i\uparrow} \rangle_{0} \langle n_{i\downarrow} \rangle_{0}
+ \dfrac{-\langle \tilde{O}^{\dagger}_{i} O_{i} \rangle_{0} 
- \langle O_{i} \tilde{O}_{i} \rangle_{0} 
+ \langle \tilde{O}^{\dagger}_{i} O_{i} \tilde{O}_{i} \rangle_{0}
+ \sum_{\sigma} \langle n_{i-\sigma} \rangle_{0} 
\langle \tilde{O}^{\dagger}_{i} \tilde{n}_{i\sigma} \tilde{O}_{i} \rangle_{0}
}
{1+\langle \tilde{O}^{\dagger}_{i} \tilde{O}_{i} \rangle_{0}} \ .
\label{dble}
\end{eqnarray}
Here $\tilde{n}_{i} = n_{i} -  \langle n_{i} \rangle_{0}$,
$\tilde{n}_{k\sigma} = n_{k\sigma} -  \langle n_{k\sigma} \rangle_{0}$, and
\begin{align} 
&\hspace{-8mm} \langle \tilde{O}^{\dagger}_{i}\tilde n_{i\sigma}\tilde{O}_{i} \rangle_{0} \nonumber \\
&\hspace{-8mm}=  U^{2} \tilde \eta^{2}
\int \frac{\bigg[ \prod \limits^5_{n=1} d\epsilon_{n}\bigg] 
\bigg[ \prod \limits^5_{n=1} \rho{(\epsilon_n)}\bigg] 
f(\tilde{\epsilon}_{1-\sigma}) 
[1-f(\tilde{\epsilon}_{2-\sigma})] 
f(\tilde{\epsilon}_{3\sigma})
[1-f(\tilde{\epsilon}_{4\sigma})]}
{(\epsilon_{4} - \epsilon_{3} + \epsilon_{2} - \epsilon_{1} -\epsilon_c)^{2}} \nonumber \\
&\hspace{10mm}\times \Big[\frac{[1-f(\tilde{\epsilon}_{5\sigma})]}
{(\epsilon_{5} - \epsilon_{3} + \epsilon_{2} - \epsilon_{1} -\epsilon_c)}
-\frac{f(\tilde{\epsilon}_{5\sigma})}
{(\epsilon_{4} - \epsilon_{5} + \epsilon_{2} - \epsilon_{1} -\epsilon_c)} \Big ].
\label{ono2}
\end{align}
%
%
%\begin{eqnarray}
%\end{eqnarray}
\begin{align} 
&\hspace{-8mm} {N} \langle \tilde{O}^{\dagger}_{i}\tilde n_{k\sigma}\tilde{O}_{i} \rangle_{0} \nonumber \\
&\hspace{-8mm}=  U^{2} \tilde \eta^{2}\bigg[[1-f(\tilde{\epsilon}_{k\sigma})]
\int\frac{\bigg[ \prod \limits^3_{n=1} d\epsilon_{n}\bigg] 
\bigg[ \prod \limits^3_{n=1} \rho{(\epsilon_n)}\bigg] 
f(\tilde{\epsilon}_{1-\sigma}) 
[1-f(\tilde{\epsilon}_{2-\sigma})] 
f(\tilde{\epsilon}_{3\sigma})}
{(\epsilon_{k\sigma} - \epsilon_{3} + \epsilon_{2} - \epsilon_{1} -\epsilon_c)^{2}} \nonumber \\
&-f(\tilde{\epsilon}_{k\sigma})
\int \frac{ \bigg[ \prod \limits^3_{n=1} d\epsilon_{n}\bigg] 
\bigg[ \prod \limits^3_{n=1} \rho{(\epsilon_n)}\bigg] 
f(\tilde{\epsilon}_{1-\sigma}) 
[1-f(\tilde{\epsilon}_{2-\sigma})] 
f(\tilde{\epsilon}_{3\sigma})}
{( \epsilon_{3}-\epsilon_{k\sigma}  + \epsilon_{2} - \epsilon_{1} - \epsilon_c)^{2}}\bigg].
\label{onko2}
\end{align}
\begin{align} 
& \langle \tilde{O}^{\dagger}_{i} O_{i} \rangle_{0} 
+ \langle O_{i} \tilde{O}_{i} \rangle_{0} \nonumber \\
&=  2U \tilde \eta
\int \frac{\bigg[ \prod \limits^4_{n=1} d\epsilon_{n}\bigg] 
\bigg[ \prod \limits^4_{n=1} \rho{(\epsilon_n)}\bigg] 
f(\tilde{\epsilon}_{1\uparrow}) 
[1-f(\tilde{\epsilon}_{2\uparrow})] 
f(\tilde{\epsilon}_{3\downarrow})
[1-f(\tilde{\epsilon}_{4\downarrow})]}
{(\epsilon_{4} - \epsilon_{3} + \epsilon_{2} - \epsilon_{1} -\epsilon_c)}.
\label{otildeo2}
\end{align}

Note that $\langle \tilde{O}^{\dagger}_{i} O_{i} \tilde{O}_{i} \rangle_{0}$ has been given by eq. (\ref{ooo02}). The expressions of these physical quantities
consist of the multiple integrals up to the 6-folds.  One can reduce
these integrals up to the 2-folds using the Laplace transform~\cite{schwe91}.  Their
expressions are given in Appendix.

\section{Numerical Results}
We have performed the numerical calculations for the non-half-filled as well as half-filled 
bands of the Hubbard model in order to examine  the validity of the improved scheme of the MLA and 
the effect of electron correlations in  the local ansatz. To calculate various physical quantities, 
we have adopted the hypercubic lattice in  infinite dimensions, where the single-site approximation works best. The density of states (DOS) for non-interacting system is given by 
$\rho(\epsilon) = (1/\sqrt{\pi}) \exp (-\epsilon^{2})$ in this case~\cite{kakeh08}. 
The energy unit is chosen to be $\int d\epsilon \rho (\epsilon)
\epsilon^{2} = 1/2$. The external magnetic field $h$ is assumed to be zero.

\subsection {Role of the best choice of  $\tilde{\eta} $ }

To calculate  various quantities in the MLA, we solved the self-consistent equations (\ref{ec})
and (\ref{etatil}) with use of the Laplace transforms of elements, which are given in Appendix.
In this sub-section, we compare the self-consistent results with the non self-consistent ones to clarify the role of the best $\tilde\eta$.

\begin{figure}
\begin{center}
\includegraphics[scale=0.50, angle=-90]{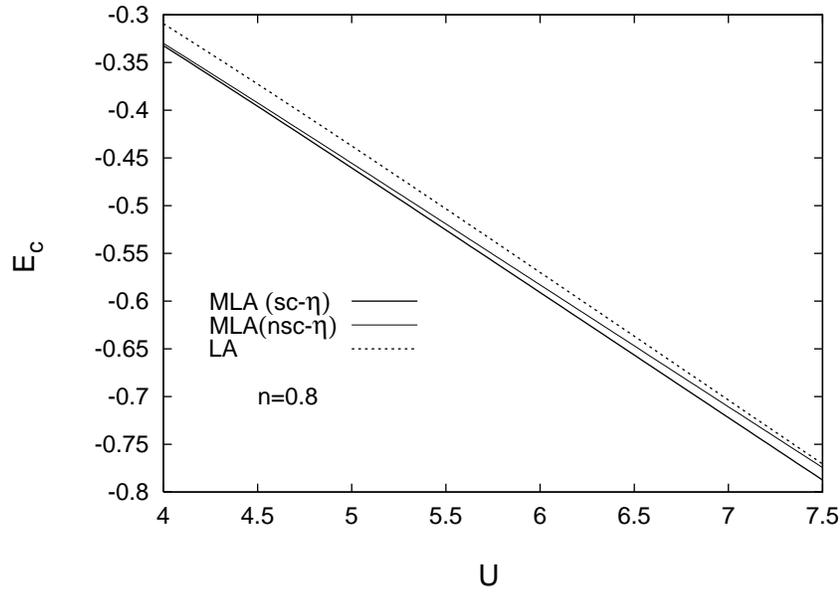}%
\caption{\label{figecn0.8}
The correlation energy $E_c$ vs. Coulomb interaction energy curve for $n=0.8$.
The thick solid curve: the MLA with self-consistent $\tilde \eta $, the thin curve: the MLA with non 
self-consistent $\tilde \eta$, and  the dashed curve: the LA.}
\end{center}
\end{figure}

Figure \ref{figecn0.8} shows the calculated correlation energy as a function 
of Coulomb interaction. The correlation energy for the  MLA without 
best choice of $\tilde\eta$ gives the lower correlation energy as compared with the LA.
The correlation energy for the MLA with the best choice of 
$\tilde\eta$ is lower than that of the non-self-consistent MLA. The results 
indicate that the self-consistency of $\tilde\eta$ is significant for finding the best energy.

\begin{figure}
\begin{center}
\includegraphics[scale=0.50, angle=-90]{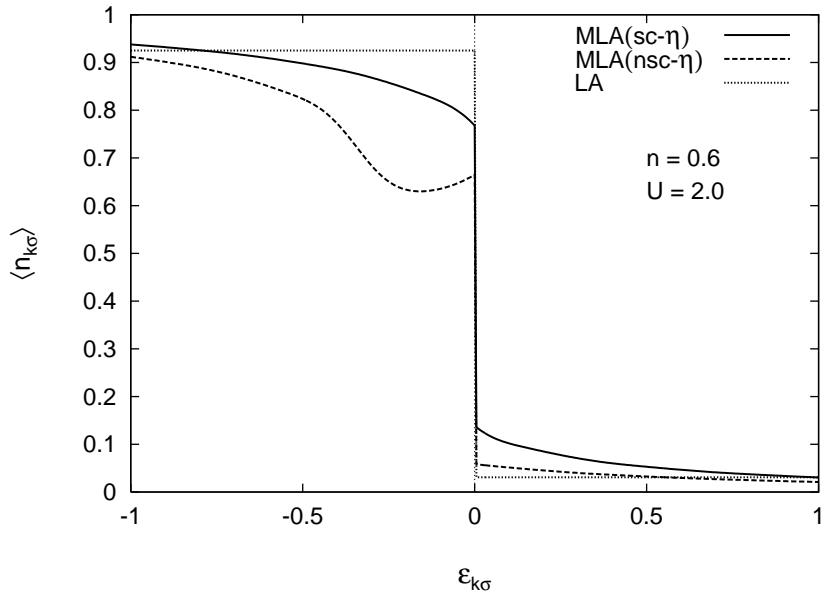}%
\caption{\label{fignku2n0.6}
The momentum distribution as a function of energy $\epsilon_{k}$ for various theories
with $n=0.6$ and $U=2.0$.
The solid curve: the MLA with the best choice of $\tilde \eta $, the 
dashed curve: the MLA without the best choice of $\tilde \eta$,  
and the dotted curve: the LA.}
\end{center}
\end{figure}

In Fig. \ref{fignku2n0.6} we show an example of the momentum distribution 
as a function of energy $\epsilon_{k\sigma}$ when electron number is deviated from 1. The MLA with non self-consistent 
$\tilde \eta $ (\ref{etatil0}) shows a bump in the vicinity of the Fermi level, leading to an unphysical result.
The MLA with self-consistent $\tilde \eta $ yields a significant 
momentum dependence  which shows monotonical  decrease of the distribution with
increasing $\epsilon_{k\sigma}$. 

\begin{figure}
\begin{center}
\includegraphics[scale=0.50, angle=-90]{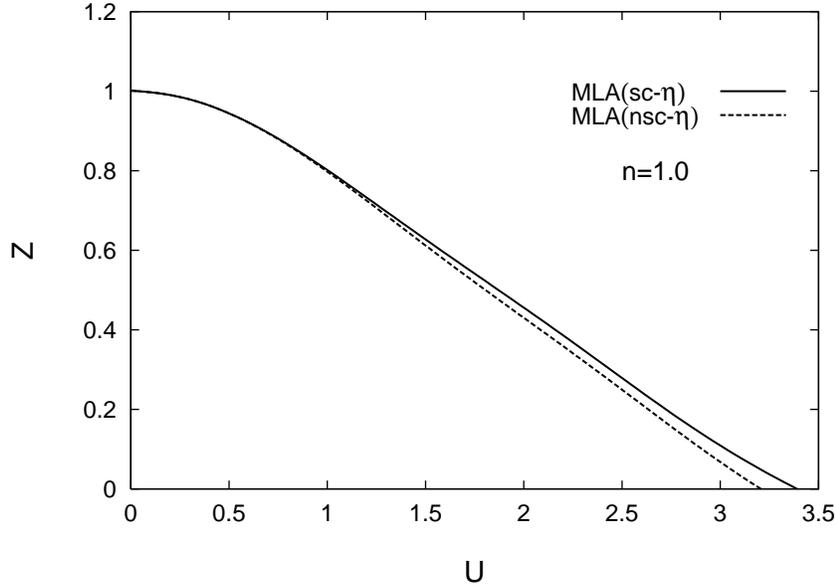}%
\caption{\label{figzu2}
Quasiparticle-weight Z vs. Coulomb interaction curves in the MLA 
with self-consistent $\tilde \eta $ (solid curve), and without (dashed curve).}
\end{center}
\end{figure}

We have also calculated   the quasiparticle weight $Z$ vs. Coulomb 
interaction energy curves at half-filling. As shown in Fig. \ref{figzu2},
we find that the best choice of  $\tilde \eta $ 
increases $Z$ ($i.e.$, decreases the inverse effective mass), so that the critical Coulomb interaction of the divergence of the effective mass, $U_{\rm c2}$  changes from $3.21$ to  $3.40$. The latter is 
closer to the NRG~\cite{bulla99} value $U_{\rm c2}$ = $4.10$, which is believed to be 
the best at present.

From the above discussions on the results with and without self-consistent $\tilde\eta$,
it is obvious  that the best choice of $\tilde\eta$ improves the results of the MLA.
In  the  following discussions we adopt the best choice of $\tilde\eta$. 

\subsection{MLA vs LA in various physical quantities}

In this section, we present the numerical results on various 
physical quantities, and discuss the new aspects of the MLA and related 
electron correlation effects by comparing the MLA with the  LA. 

\begin{figure}[htbp]
\begin{center}
\includegraphics[scale=0.50, angle=-90]{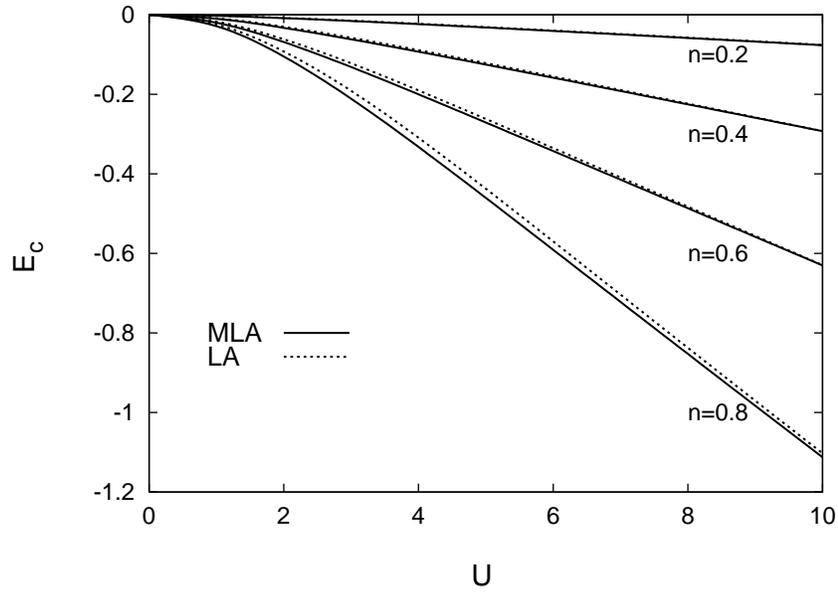} 
\caption{\label{figec}
The correlation energies $E_c$ vs. Coulomb interaction energy parameter $U$ 
in the MLA (solid curve) and the LA (dashed curve) for various electron number $n$.}
\end{center}
\end{figure}

In Fig. \ref{figec}, we represent the  calculated correlation energy per atom as a 
function of Coulomb interaction $ U $. The energy in the MLA is lower than that of the LA
over all Coulomb interaction energy parameters $U$ and electron numbers $n$. These results imply that  the MLA improves the LA. The magnitude of the correlation energy $|\epsilon_c|$ tends to increase with increasing $U$, because with increasing $U$ the correlation corrections increase as $U^2$ for small $U$ and cancel the Hartree-Fock energy loss being linear in $U$ for large $U$.
For a fixed value of the Coulomb interaction $U$, the gain of the  correlation energy 
$|\epsilon_c|$ increases with increasing $n$, because there is a correlation energy 
gain at each doubly-occupied site and the number of such sites increases with increasing $n$.

\begin{figure}[htbp]
\begin{center}
\includegraphics [scale=0.50, angle=-90]{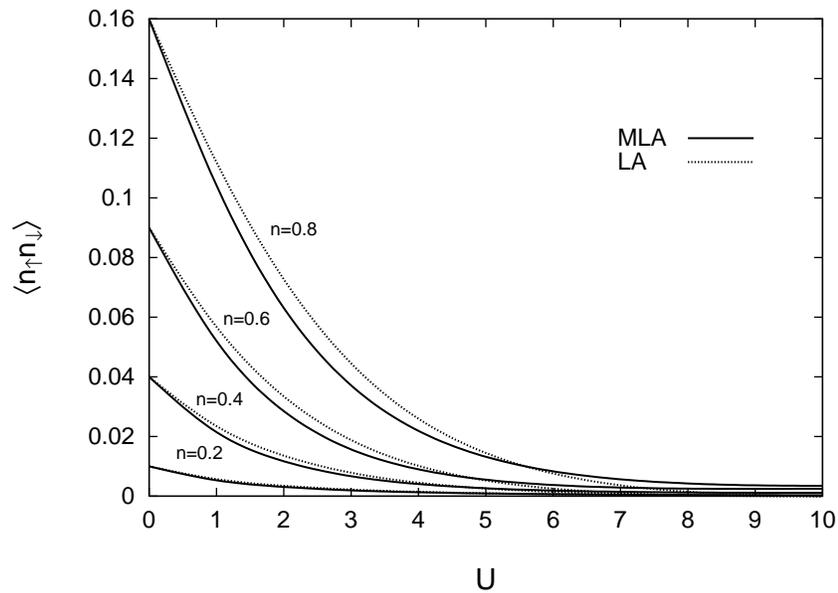}%
\caption{\label{figdbl}
The double occupation number $\langle n_{\uparrow}n_{\downarrow} \rangle$ 
vs. Coulomb interaction energy $U$ curves 
in the MLA (solid curve) and the LA (dotted curve).}
\end{center}
\end{figure}

Figure \ref{figdbl} depicts the double occupation number
$\langle n_{\uparrow}n_{\downarrow} \rangle$ vs. Coulomb 
interaction curves for the non-half-filled case.
In the uncorrelated limit, the double occupancy  is the same for both LA and MLA
and it decreases with increasing Coulomb interaction $U$ because 
electrons move on the lattice so as to suppress the loss of Coulomb energy due to double 
occupation. We find that the MLA wavefunction reduces the 
double occupancy as compared with that of the LA in the range $0<U\lesssim5$, while in the range 
$5\lesssim U$ the double occupancy in the MLA is larger than that of the LA. 
It implies that the LA  with momentum-independent $\eta_{LA}$ overestimates the itinerant
character for weak and intermediate $U$ regions, while it overestimates the atomic character for large $U$ region.

\begin{figure}[htbp]
\begin{center}
\includegraphics[scale=0.50, angle=-90]{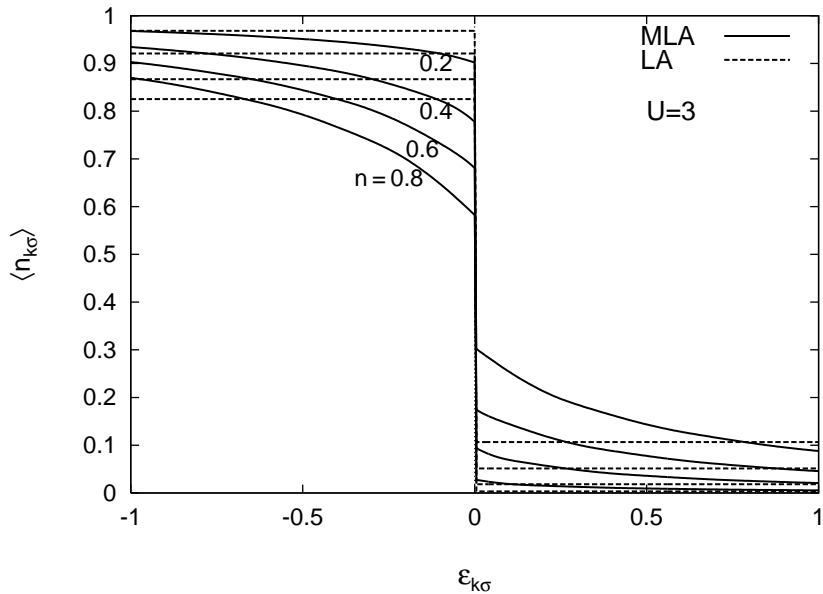} 
\caption{\label{fignk}
The momentum distribution as a function of energy $\epsilon_{k}$ 
for various electron number with constant  Coulomb interaction energy 
parameters $U=3$. The MLA: solid curves, the LA: dashed curves.}
\end{center}
\end{figure}

The momentum-distribution function shown   in Fig. \ref{fignk}  indicates more 
distinct difference between the LA and the MLA. The  distributions in the LA are
constant below and above the Fermi level irrespective of $U$. The same behavior 
is also found in the GA~\cite{gutz63,gutz64,gutz65}. The MLA curves show a monotonical decrease of the 
distribution with increasing $\epsilon_{k\sigma}$, indicating  a distinct 
momentum dependence  of $\langle n_{k\sigma} \rangle$ via energy $\epsilon_{k\sigma}$,
which is qualitatively different from both the LA  and the GA.

\begin{figure}[htbp]
\begin{center}
\includegraphics[scale=0.50, angle=-90]{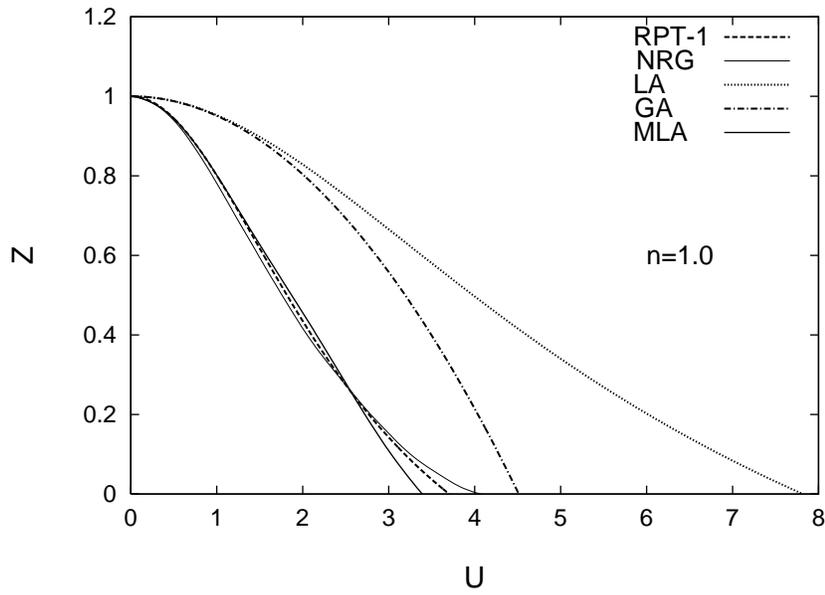}%
\caption{\label{figzu1}
Quasiparticle-weight vs. Coulomb interaction curves in various theories.
The RPT-1: dashed curve~\cite{kakeh04}, the NRG: thin solid curve~\cite{bulla99}, 
the LA: dotted curve, the MLA: solid curve, and the GA: dot-dashed 
curve. }
\end{center}
\end{figure}

The quasiparticle weight $Z$ ({\it i.e.} the inverse effective mass) is obtained from 
the jump at the Fermi level in the momentum distribution according to the Fermi liquid theory~\cite{landu57,gbaym91}.
Calculated quasiparticle weight vs Coulomb interaction curves are shown in Fig. \ref{figzu1}
for the half-filled case. The  quasiparticle weight in the LA changes as
$Z=(1-3\eta^{2}_{\rm \, LA}/16)/(1+\eta^{2}_{\rm \, LA}/16)$ and
vanishes at $U_{\rm c2}(\rm LA) = 24/\sqrt{3\pi} \, (=7.82)$.  
In the GA~\cite{br70}, the quasiparticle weight changes as
$Z=1-(U/U_{\rm c2})^{2}$.  The curve in the GA agrees with the LA curve for
small $U$.  But it deviates from the LA when $U$ becomes larger, 
and vanishes at $U_{\rm c2}(\rm GA)=8/\sqrt{\pi} \, (=4.51)$.  
It should be noted that the GA curve strongly deviates from the curve 
in the NRG~\cite{bulla99} which is considered to be the best.  
We observe that the critical Coulomb interaction $U_{\rm c2}$ for the 
self-consistent $\tilde\eta $ is $3.40$  in the MLA while $U_{\rm c2}$ in  the 
non-self-consistent $\tilde\eta $  yields $3.21$.
The quasiparticle weight in the MLA much improves the LA as seen in Fig.\ref{figzu1}. 
We note that the wavefunction itself
does not show the metal-insulator transition at $U_{\rm c2}$ in the present 
approximation because the approximate
expression of variational parameters (\ref{etaint})
has no singularity at finite value of $U$.
In this sense, the calculated
$Z$ and wavefunction are not self-consistent in the present
approximation.  The values of $Z$ obtained by the LA and the MLA should be regarded
as an estimate from the metallic side. 

\begin{figure}[htbp]
\begin{center}
\includegraphics[scale=0.50, angle=-90]{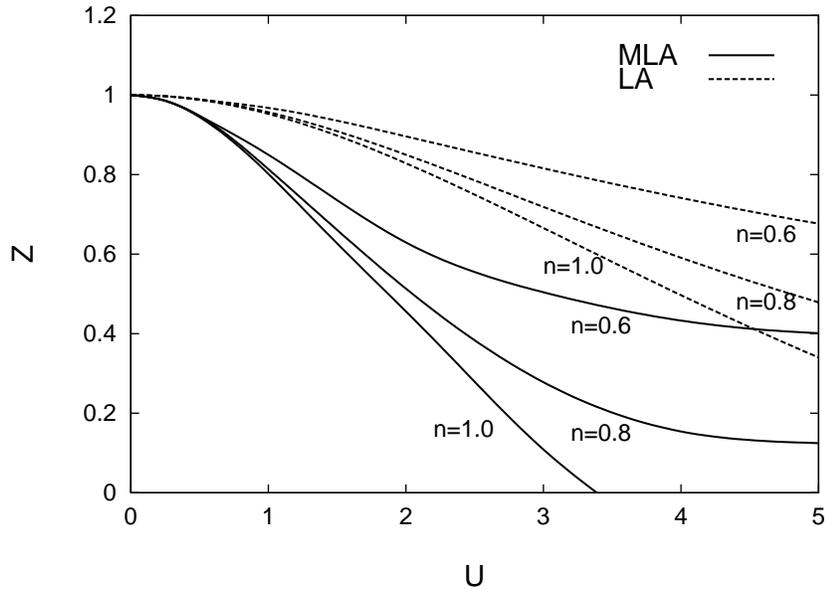}%
\caption{\label{figzu}
Quasiparticle-weight Z vs. Coulomb interaction curves.
The MLA: solid curve, the LA: dashed curve.}
\end{center}
\end{figure}

Figure \ref{figzu} also shows the quasiparticle weight as a function of
Coulomb interaction energy $U$ for the non-half-filled case and the half-filled case. 
In  the uncorrelated limit, the quasiparticle weight is $1$  as it should be. 
It decreases with increasing Coulomb interaction $U$ for both the MLA and the
LA. In general, the curves in the MLA are lower than those in the LA as expected 
from the fact  $U_{c2} (\rm MLA)$ $<U_{c2}(\rm LA) $ at half-filling. This means 
that the electron effective mass is enhanced by the self-consistent treatment of 
$\tilde{\eta}$ irrespective of $U$ and $n$.

\section{Summary and Discussions}

In the present paper, we improved the variational scheme of the MLA which
self-consistently determines both the variational amplitude $\tilde\eta$ 
and the correlation energy $\epsilon_c$  making use of variational 
principles. To examine  the improvement and  validity of the theory, we performed the  numerical calculations  on the basis of the Hubbard model on the hypercubic lattice in  infinite dimensions. 
We verified for both the half-filled  and the non half-filled  bands that the self-consistent scheme of the MLA improves the correlation energy,
the momentum distribution function as well as the quasiparticle weight.
Therefore, the self-consistency of  $\tilde\eta$ is 
significant for quantitative understanding of electron correlations.

Within the self-consistent MLA, we have clarified the role of the 
momentum dependence of variational parameters  in comparison 
with the original LA. We demonstrated that the self-consistent
MLA improves the LA irrespective of the Coulomb interaction energy 
parameter $U$ and electron number $n$. The correlation energy in 
the MLA is  lower than those of the LA  and the GA in the weak 
and intermediate Coulomb interaction regimes. Thus the MLA 
wavefunction should be better than both the LA and the GA in these regimes. 
The double occupation number is suppressed as compared with the 
LA both in the same interaction regimes.
We found that the calculated momentum distribution functions show a distinct 
momentum dependence. This is qualitatively different from the LA and the GA because
both of them lead to the momentum-independence of the distributions below and
above the Fermi level. We also found that the quasiparticle weight in 
the MLA is lower than that of the LA irrespective of $U$ and $n$, 
 and is close to the result of the NRG. Accordingly  the critical Coulomb interaction $U_{c2}$ 
of the MLA becomes closer to that obtained in the NRG. 

The variational theories discussed in the present paper construct the correlated 
ground state by applying the two-particle operators to the Hartree-Fock ground state. 
The wavefunctions of  both  the LA and the GA are expressed by the momentum dependent 
two-particle excited states in addition to the Hartree-Fock one. Each amplitude of the 
excited states is momentum independent in these methods. The MLA wave function, on the
other hand, each amplitude of the two-particle excited states is momentum dependent.
By choosing  the momentum-dependent amplitudes  $ \eta_{k^{\prime}_{2}k_{2}k^{\prime}_{1}k_{1}}$
best on the basis  of the variational principle, we improved the LA and the GA in the weak and 
intermediate Coulomb interaction regimes. Note that this procedure does not depend on 
dimensions of the system within the SSA because of the local projection 
$\langle k^{\prime}_{1}|i \rangle \langle i|k_{1} \rangle 
\langle k^{\prime}_{2}|i \rangle \langle i|k_{2} \rangle$ in the operator $\tilde{O_i}$
and thus the improvement remains unchanged even in infinite dimensions.

Needless to say, many methods to solve the correlation problems in infinite dimensions have been 
developed~\cite{georges96,kakeh04-II}. The  NRG is one of the best approaches to calculate the excitations at zero temperature
as well as related ground-state properties. The accuracy of MLA is on the level of the iterative 
perturbation theory~\cite{georges93} at the present stage. Furthermore excitation properties cannot directly be calculated 
by the variational approach. It is also true, however, that the high-quality methods such as the 
NRG~\cite{bulla99} are not applicable to the realistic systems because of their complexity. The present approach is 
applicable to more complex systems. Moreover the wavefunction method arrows us to calculate any static averages
as we have demonstrated in paper I~\cite{kakeh08}. We therefore believe that further developments of the MLA wavefunction  
approach will provide us  with a useful tool for understanding correlated electrons in 
the realistic systems and their physics. Developments of the theory towards the strongly correlated system 
and its extension to the realistic systems are our current problems in progress.  

\section*{Acknowledgment}
The present work is supported by Grant-in-Aid for Scientific Research KAKENHI (22540395). 

\appendix

\section{Laplace transform for the correlation
 calculations} 
 
The Laplace transform can significantly reduce the number of integrals in 
the physical quantities which appear in our variational theory. It is  written as follows:
\begin{eqnarray}
\dfrac{1}{z - \epsilon_{4} + \epsilon_{3} - \epsilon_{2} + \epsilon_{1}
 + \epsilon_{c}} = -i \int^{\infty}_{0} dt \,
{\rm
e}^{i(z-\epsilon_{4}+\epsilon_{3}-\epsilon_{2}+\epsilon_{1}
+\epsilon_{\rm c}) \, t}
\ .
\label{laplace}
\end{eqnarray}
Here $z=\omega+i\delta$, and $\delta$ is an infinitesimal positive
number.

The term  $\langle \tilde{O}^{\dagger}_{i}\tilde{O}_{i} \rangle_{0}$  in eq. (\ref{oo2}) can be written  in the energy 
as follows:
\begin{align} 
& \langle \tilde{O}^{\dagger}_{i}\tilde{O}_{i} \rangle_{0} \nonumber \\
&=  U^{2} \tilde \eta^{2}
\int \frac{\bigg[ \prod \limits^4_{n=1} d\epsilon_{n}\bigg] 
\bigg[ \prod \limits^4_{n=1} \rho{(\epsilon_n)}\bigg] 
f(\tilde{\epsilon}_{1\uparrow}) 
[1-f(\tilde{\epsilon}_{2\uparrow})] 
f(\tilde{\epsilon}_{3\downarrow})
[1-f(\tilde{\epsilon}_{4\downarrow})]}
{(\epsilon_{4} - \epsilon_{3} + \epsilon_{2} - \epsilon_{1} -\epsilon_c)^{2}}
\label{oo3} \\
&=  U^{2} \tilde \eta^{2}
\lim_{z\to 0}  \int \frac{\bigg[ \prod \limits^4_{n=1} d\epsilon_{n}\bigg] 
\bigg[ \prod \limits^4_{n=1} \rho{(\epsilon_n)}\bigg] 
f(\tilde{\epsilon}_{1\uparrow}) 
[1-f(\tilde{\epsilon}_{2\uparrow})] 
f(\tilde{\epsilon}_{3\downarrow})
[1-f(\tilde{\epsilon}_{4\downarrow})]}
{(z-\epsilon_{4} - \epsilon_{3} + \epsilon_{2} - \epsilon_{1} -\epsilon_c)^{2}}.
\end{align}
Here, $\tilde\epsilon_{n\sigma}=\epsilon_n+\tilde{\epsilon}_{\sigma}$, and  $\tilde{\epsilon}_{\sigma} = \epsilon_{0} 
+ U \langle n_{i-\sigma} \rangle_{0} - \mu$ is the Hartree-Fock level
measured from the Fermi level $\mu$.  $\rho(\epsilon)$ in the above
expressions denotes the density of states for $\epsilon_{k}$, {\it i.e.},
the Fourier transform of $t_{ij}$.
Now using the relation of Laplace transform (\ref{laplace}), we obtain 
\begin{align}
\langle \tilde{O}^{\dagger}_{i}\tilde{O}_{i} \rangle_{0} & = 
-U^{2} \tilde \eta^{2}\lim_{z\to 0}
\int_0^\infty dt dt' e^{(z+\epsilon_c)(t+t')}
\int d\epsilon_1 e^{i\epsilon_1(t+t')} \rho(\epsilon_1) f(\tilde{\epsilon}_{1\uparrow})\nonumber \\
& \hspace{.5cm}\times \int d\epsilon_2 e^{-i\epsilon_2(t+t')} \rho(\epsilon_2) [1-f(\tilde{\epsilon}_{2\uparrow})]
\int d\epsilon_3 e^{i\epsilon_3(t+t')} \rho(\epsilon_3) f(\tilde{\epsilon}_{3\downarrow})\nonumber \\
& \hspace{.5cm}\times \int d\epsilon_4 e^{-i\epsilon_4(t+t')} \rho(\epsilon_4) [1-f(\tilde{\epsilon}_{4\downarrow})]\\
& = - U^{2} \tilde \eta^{2}  
\int^{\infty}_{0} \! dtdt^{\prime} 
{\rm e}^{i {\epsilon_c}(t+t^{\prime})}
a_{\uparrow}(-t-t^{\prime})b_{\uparrow}(t+t^{\prime})
a_{\downarrow}(-t-t^{\prime})b_{\downarrow}(t+t^{\prime}).
\label{loo}
\end{align}
Here
\begin{eqnarray}
a_{\sigma}(t) = \int d\epsilon \rho(\epsilon)
f(\epsilon+\tilde{\epsilon}_{\sigma}) \, {\rm e}^{-i\epsilon t}
\ ,
\label{asigma}
\end{eqnarray}
\begin{eqnarray}
b_{\sigma}(t) = \int d\epsilon \rho(\epsilon)
[1-f(\epsilon+\tilde{\epsilon}_{\sigma})] \, {\rm e}^{-i\epsilon t}
\ .
\label{bsigma}
\end{eqnarray}
The 4-fold integrals of $\langle \tilde{O}^{\dagger}_{i}\tilde{O}_{i} \rangle_{0}$ in eq. (\ref{oo3})
reduce to the 2-fold integrals in eq. (\ref{loo}).

In the same way, we can perform the Laplace transform of various elements in the physical quantities as follows:
\begin{eqnarray}
\langle \tilde{H} \tilde{O}_{i}\rangle_{0} 
& =  & \langle \tilde{O}^{\dagger}_{i}\tilde{H} \rangle_{0}^{\ast}
\nonumber \\
& = & i U^{2} \tilde \eta
\int^{\infty}_{0} \! dt \, {\rm e}^{i{ \epsilon_c}t} \, 
a_{\uparrow}(-t)a_{\downarrow}(-t)b_{\uparrow}(t)b_{\downarrow}(t) \ ,
\hspace{5mm}
\label{lho}
\end{eqnarray}
\begin{eqnarray}
\langle \tilde{O}^{\dagger}_{i} \tilde{H_0} \tilde{O}_{i} \rangle_{0} & = &
- U^{2} \tilde \eta^{2}  
\int^{\infty}_{0} \! dtdt^{\prime} 
{ e}^{i{ \epsilon_c}(t+t^{\prime})}
\big[
a_{\uparrow}(-t-t^{\prime})b_{\uparrow}(t+t^{\prime})
a_{\downarrow}(-t-t^{\prime})b_{1\downarrow}(t+t^{\prime})  \nonumber \\
& & \hspace{30mm}
-a_{\uparrow}(-t-t^{\prime})b_{\uparrow}(t+t^{\prime})
a_{1\downarrow}(-t-t^{\prime})b_{\downarrow}(t+t^{\prime})  \nonumber \\
& & \hspace{30mm}
+a_{\uparrow}(-t-t^{\prime})b_{1\uparrow}(t+t^{\prime})
a_{\downarrow}(-t-t^{\prime})b_{\downarrow}(t+t^{\prime})  \nonumber \\
& & \hspace{30mm}
-a_{1\uparrow}(-t-t^{\prime})b_{\uparrow}(t+t^{\prime})
a_{\downarrow}(-t-t^{\prime})b_{\downarrow}(t+t^{\prime})
\big] \ , \hspace{10mm}
\label{loh0o}
\end{eqnarray}
\begin{eqnarray}
\langle \tilde{O}^{\dagger}_{i} O_{i} \tilde{O}_{i} \rangle_{0} & = &
- U^{2} \tilde \eta^{2}   
\int^{\infty}_{0} \! dtdt^{\prime} 
{\rm e}^{i {\epsilon_c}(t+t^{\prime})}
\big[
a_{\uparrow}(-t)b_{\uparrow}(t+t^{\prime})
a_{\downarrow}(-t)b_{\downarrow}(t+t^{\prime})
a_{\uparrow}(-t^{\prime})a_{\downarrow}(-t^{\prime})  \nonumber \\
& & \hspace{30mm}
-a_{\uparrow}(-t)b_{\uparrow}(t+t^{\prime})
a_{\downarrow}(-t-t^{\prime})b_{\downarrow}(t)
a_{\uparrow}(-t^{\prime})b_{\downarrow}(t^{\prime})  \nonumber \\
& & \hspace{30mm}
-a_{\uparrow}(-t-t^{\prime})b_{\uparrow}(t)
a_{\downarrow}(-t)b_{\downarrow}(t+t^{\prime})
b_{\uparrow}(t^{\prime})a_{\downarrow}(-t^{\prime})  \nonumber \\
& & \hspace{30mm}
+a_{\uparrow}(-t-t^{\prime})b_{\uparrow}(t)
a_{\downarrow}(-t-t^{\prime})b_{\downarrow}(t)
b_{\uparrow}(t^{\prime})b_{\downarrow}(t^{\prime})
\big] \ . \hspace{10mm}
\label{looo}
\end{eqnarray}
Here
\begin{eqnarray}
a_{1\sigma}(t) = \int d\epsilon \rho(\epsilon)
f(\epsilon+\tilde{\epsilon}_{\sigma}) \, \epsilon \, { e}^{-i\epsilon t}
\ ,
\label{a1sigma}
\end{eqnarray}
\begin{eqnarray}
b_{1\sigma}(t) = \int d\epsilon \rho(\epsilon)
[(1-f(\epsilon+\tilde{\epsilon}_{\sigma})] \, \epsilon \, {e}^{-i\epsilon t}
\ .
\label{b1sigma}
\end{eqnarray}

The element (\ref{etatlnu}) and (\ref{etatlde}) for the calculation of 
the best choose of $\tilde{\eta}$ are expressed as
\begin{eqnarray}
C &= &  -
\int^{\infty}_{0} \! dtdt^{\prime} 
{ e}^{i {\epsilon_c}(t+t^{\prime})}
\big[
a_{\uparrow}(-t)b_{\uparrow}(t+t^{\prime})
a_{\downarrow}(-t)b_{\downarrow}(t+t^{\prime})
a_{\uparrow}(-t^{\prime})a_{\downarrow}(-t^{\prime})  \nonumber \\
& & \hspace{17mm}
-a_{\uparrow}(-t-t^{\prime})b_{\uparrow}(t)
a_{\downarrow}(-t)b_{\downarrow}(t+t^{\prime})
b_{\uparrow}(t^{\prime})a_{\downarrow}(-t^{\prime})  \nonumber \\
& & \hspace{17mm}
-a_{\uparrow}(-t)b_{\uparrow}(t+t^{\prime})
a_{\downarrow}(-t-t^{\prime})b_{\downarrow}(t)
a_{\uparrow}(-t^{\prime})b_{\downarrow}(t^{\prime})  \nonumber \\
& & \hspace{17mm}
+a_{\uparrow}(-t-t^{\prime})b_{\uparrow}(t)
a_{\downarrow}(-t-t^{\prime})b_{\downarrow}(t)
b_{\uparrow}(t^{\prime})b_{\downarrow}(t^{\prime})
\big] \ , \hspace{10mm}
\label{letatlde}
\end{eqnarray}
and
\begin{eqnarray}
D =  i \int^{\infty}_{0} \! dt \, { e}^{i{ \epsilon_c}t} \, 
a_{\uparrow}(-t)a_{\downarrow}(-t)b_{\uparrow}(t)b_{\downarrow}(t) \ .
\hspace{5mm}
\label{letatlnu}
\end{eqnarray}
The correlation contribution to the momentum distribution function
(\ref{onko2}) is given by
\begin{eqnarray}
N \langle \tilde{O}^{\dagger}_{i} \tilde{n}_{k\sigma} \tilde{O}_{i} \rangle_{0} 
& = & 
U^{2} \tilde \eta^{2}  
\int^{\infty}_{0} \! dtdt^{\prime} 
{e}^{i {\epsilon_c}(t+t^{\prime})}
a_{-\sigma}(-t-t^{\prime})b_{-\sigma}(t+t^{\prime}) \nonumber \\
& &  \hspace{-15mm}  \times
\big[ f(\tilde\epsilon_{k\sigma}) b_{\sigma}(t+t^{\prime}) 
{e}^{i \epsilon_k (t+t^{\prime})}
-[1 - f(\tilde\epsilon_{k\sigma})] a_{\sigma}(-t-t^{\prime})
{\rm e}^{-i\epsilon_k (t+t^{\prime})}
\big]
\ .
\label{lonko}
\end{eqnarray}
The correlation contribution to the electron number (\ref{ono2}) 
which appears in the
calculation of the double occupation number is expressed as
\begin{eqnarray}
\hspace{-8mm}
\langle \tilde{O}^{\dagger}_{i} \tilde{n}_{i\sigma} \tilde{O}_{i} \rangle_{0} 
& = & 
- U^{2} \tilde\eta^{2}   
\int^{\infty}_{0} \! dtdt^{\prime} 
{\rm e}^{i {\epsilon_c}(t+t^{\prime})}
\big[
a_{-\sigma}(-t-t^{\prime})b_{-\sigma}(t+t^{\prime})
a_{\sigma}(-t-t^{\prime})b_{\sigma}(t)b_{\sigma}(t^{\prime}) \nonumber \\
& &  \hspace{20mm}
-a_{-\sigma}(-t-t^{\prime})b_{-\sigma}(t+t^{\prime})
a_{\sigma}(-t)b_{\sigma}(t+t^{\prime})a_{\sigma}(t^{\prime})
\big] \ .  \hspace{10mm}
\label{lonio}
\end{eqnarray}
The element (\ref{otildeo2}) for the calculation of the double occupancy
is expressed as
\begin{eqnarray}
\langle \tilde{O}^{\dagger}_{i} O_{i} \rangle_{0} 
+ \langle O_{i} \tilde{O}_{i} \rangle_{0} & = &
2iU \tilde \eta
\int^{\infty}_{0} \! dt \,
{\rm e}^{i {\epsilon_c} t}
a_{\uparrow}(-t)b_{\uparrow}(t)
a_{\downarrow}(-t)b_{\downarrow}(t) \ . 
\hspace{15mm}
\label{lotildeo}
\end{eqnarray}

\end{document}